\documentclass[10pt]{article}

\usepackage{cite,color}
\usepackage{graphicx}
\usepackage{subfigure}
\usepackage{mathptmx}
\usepackage{helvet}
\usepackage{verbatim}
\usepackage{jae_margins}

\begin{document}

\def\xspace{\enskip}
\def\antibar#1{\ensuremath{#1\bar{#1}}}
\def\tbar{\ensuremath{\bar{t}}}
\def\cbar{\ensuremath{\bar{c}}}
\def\ccbar{\antibar{c}}
\def\bbar{\ensuremath{\bar{b}}}
\def\bbbar{\antibar{b}}
\def\ttbar{\antibar{t}}
\def\ttH{\antibar{t}H}
\def\to{\ensuremath{\rightarrow}}


\title{Sensitivity of LHC experiments to the $\ttH$ final state, with $H \to \bbbar$, at center of mass energy of 14 TeV} 
\author {Ricardo Goncalo$^a$, Stefan Guindon$^b$, Vivek Jain$^b$ \\ {\it $^a$Royal Holloway, University of London, $^b$State University of New York, Albany}}
\maketitle

\begin{abstract}

With the discovery of a Higgs-like particle in 2012, attention has now turned to measuring its properties, e.g., coupling to various bosonic and fermionic final states, its spin and parity, etc. In this note, we study the sensitivity of experiments at the LHC to its coupling to the top quark, by searching for the process pp $\rightarrow \ttH$, where the primary decay mode of the $H$ is $\to \bbbar$. In this paper, the $\ttbar$ system is detected in the dilepton final state. This study is performed assuming a center of mass energy of 14 TeV and integrated luminosities of 300 fb$^{-1}$ (with an average pileup ($\mu$) of 50 additional collisions per bunch crossing) and 3000 fb$^{-1}$ (with $\mu=140$). We include systematic uncertainties in production cross-sections as well as the more important experimental uncertainties. Preliminary studies indicate that we will observe the $\ttH$ final state with a significance of 2.4 and $\ge 5.3$ for the two luminosity scenarios, respectively; addition of other $\ttbar$ final states should increase the overall significance for observing $\ttH$.
\end{abstract}

\bigskip

\section{Introduction}
\label{sect:intro}

\begin{figure}[htbp]
\begin{center}
\includegraphics[width=0.4\linewidth]{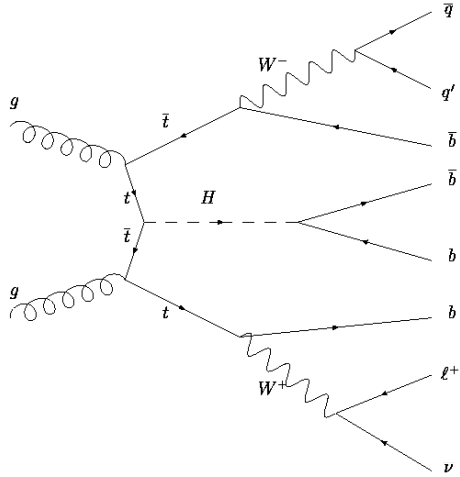}
\caption{{\bf Feynman diagram representing the lepton+jets final state of $t\bar{t}H$ production.}}
\label{fig:ttHFeyn}
\end{center}
\end{figure}

The large top quark mass implies that its coupling to the Higgs boson will be very large. Since the top quark is heavier than the Higgs-like particle, the decay $H \to \ttbar$ is kinematically suppressed. To gain direct access to the top-quark Yukawa coupling, one has to use processes where the Higgs boson is produced in association with a top quark pair. An example is shown in Fig.~\ref{fig:ttHFeyn}. Here the Higgs boson is produced by the fusion of virtual top quark pairs . It could also be  ``radiated'' off one of the top quark lines. Given the mass of the Higgs boson, the dominant decay mode in the Standard Model is to the $\bbbar$ final state.

In this study, we reconstruct the $\ttH$ final state by reconstructing the $\ttbar$ system in the dilepton final state (where both W bosons from top quarks decay leptonically)\footnote{One can also use the lepton+jets final state (i.e., one W from the top quark decays hadronically and the other leptonically), as shown in Fig.~\ref{fig:ttHFeyn}, and the final state where both W bosons decay hadronically.}, and the Higgs in the di-jet final state, where the primary component will be $H \to \bbbar$. Other Higgs decay modes, such as $\ttH$ production with $H \to WW$ decays and other sub-leading modes, also contribute to the signal, but we do not explicitly reconstruct them. The background from various sources, e.g., $\ttbar$ (plus an additional $\bbbar$ or light jet pair), W/Z+jets, etc., is significantly larger than the signal cross-section, so to maximize the sensitivity of the search, we need to have a very good understanding of systematic uncertainties, both experimental and theoretical. The former includes contributions such as b-tagging, and jet energy scale, whereas the latter includes understanding the cross-section and kinematics for $\ttbar$ (plus additional jets).

At a center of mass energy of 14 TeV and for a Higgs boson mass of $m_H = 125$ GeV), the signal production cross section is $\sigma(\ttH \!) = 0.6^{+14.8\%}_{-18.2\%}$ pb~\cite{higgsdecay1,higgsdecay2,higgsdecay3,higgsdecay4}, whereas the cross section of the dominant background is much larger: $\sigma(\ttbar \!)$ = 980 pb.

\section{Samples and Event Selection}
\label{sect:events}

The signal, $\ttH$, and the backgrounds, are simulated using the Delphes 3.0.9 simulation framework~\cite{delphes}. The events were generated using MadGraph~\cite{madgraph}. Samples have been generated at center of mass energy of 14 TeV. Signal samples were generated with inclusive Higgs and top quark decays. The $\ttH$ event topology is defined by the decay products of the two top quarks ($t \rightarrow b W$) and the Higgs.  In this analysis, the selection of Higgs decays to two $b$ quarks is favored.  To reduce the effect of pile-up due to extra jets and the relatively large mis-match of jet-ordering, the $W$ bosons are required to decay leptonically into a charged lepton (e/$\mu$) and neutrino (the contribution from $W \rightarrow \tau \nu$, where the
$\tau \rightarrow e/\mu \nu {\bar {\nu}}$ is included by default).  Although this top decay channel has the smallest branching ratio ($< 10$ \% of all top pair decays), however, it is the cleanest signature, with smaller background contamination relative to other channels.

The final signature contains 4 jets, all of which are $b$ jets, as well as two oppositely charged leptons and missing energy due to the two escaping neutrinos.  An event is required to have exactly two oppositely-charged leptons with transverse momentum $p_T > 15$ GeV and pseudorapidity $|\eta| < 2.5$, as well as at least 2 jets with $p_T > 25$ GeV and $|\eta| < 2.5$.  To select a sample enriched in  b-jets, a loose and a tight b-tagger are employed, approximating the performance of realistic tight (less efficient but more pure) and loose (more efficient but less pure) tagging algorithms.  To exclude background contamination from Electroweak processes, we veto events containing leptons of identical flavors with invariant mass $m_{\ell\ell}$ within 8 GeV of the nominal $Z$ mass.  In addition, $m_{\ell\ell}$ is required to be $> 15$ GeV.

\section{Basic Analysis Strategy}
\label{sect:fit}

\begin{figure}[htbp]
\begin{center}
\includegraphics[width=0.48\linewidth]{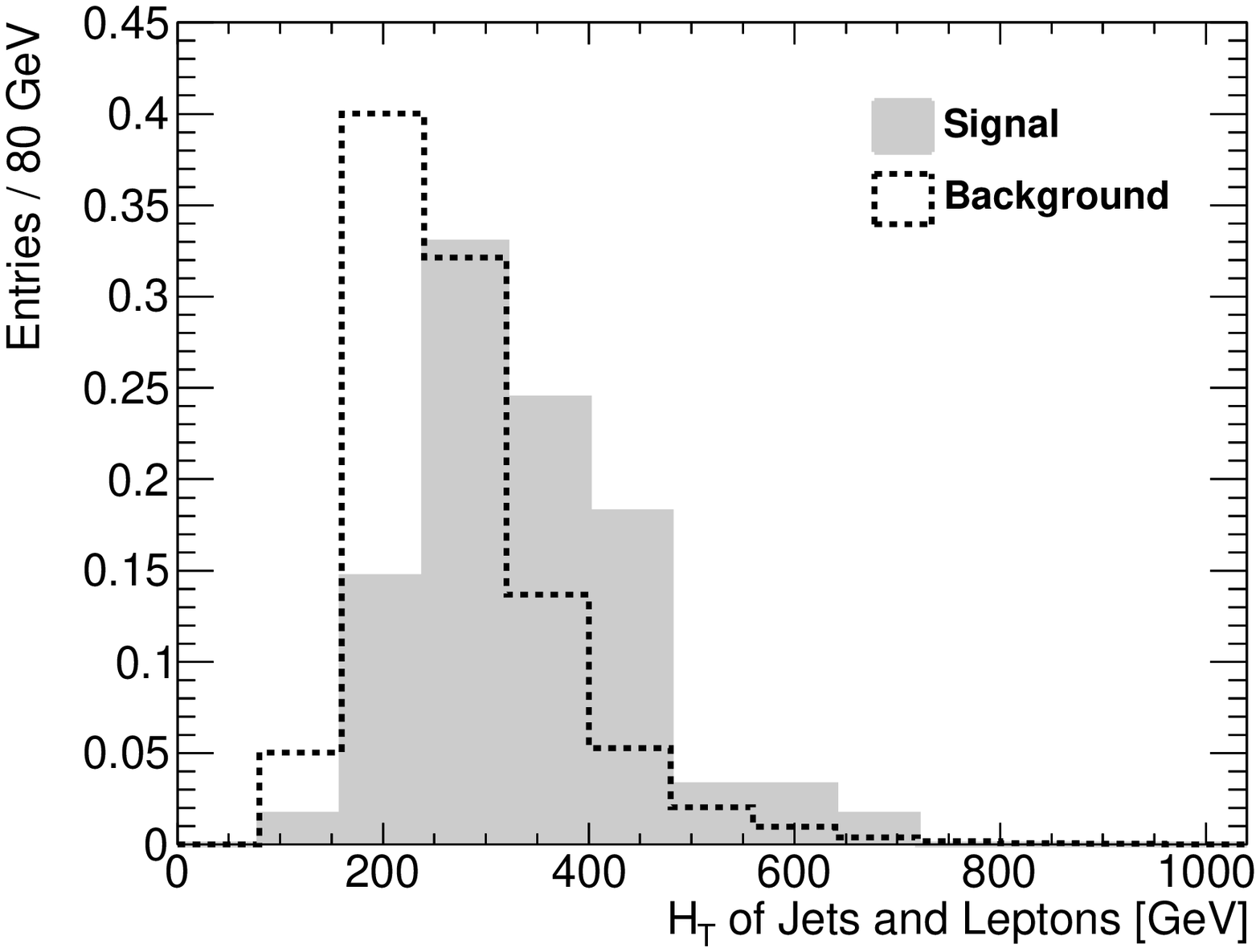}
\includegraphics[width=0.48\linewidth]{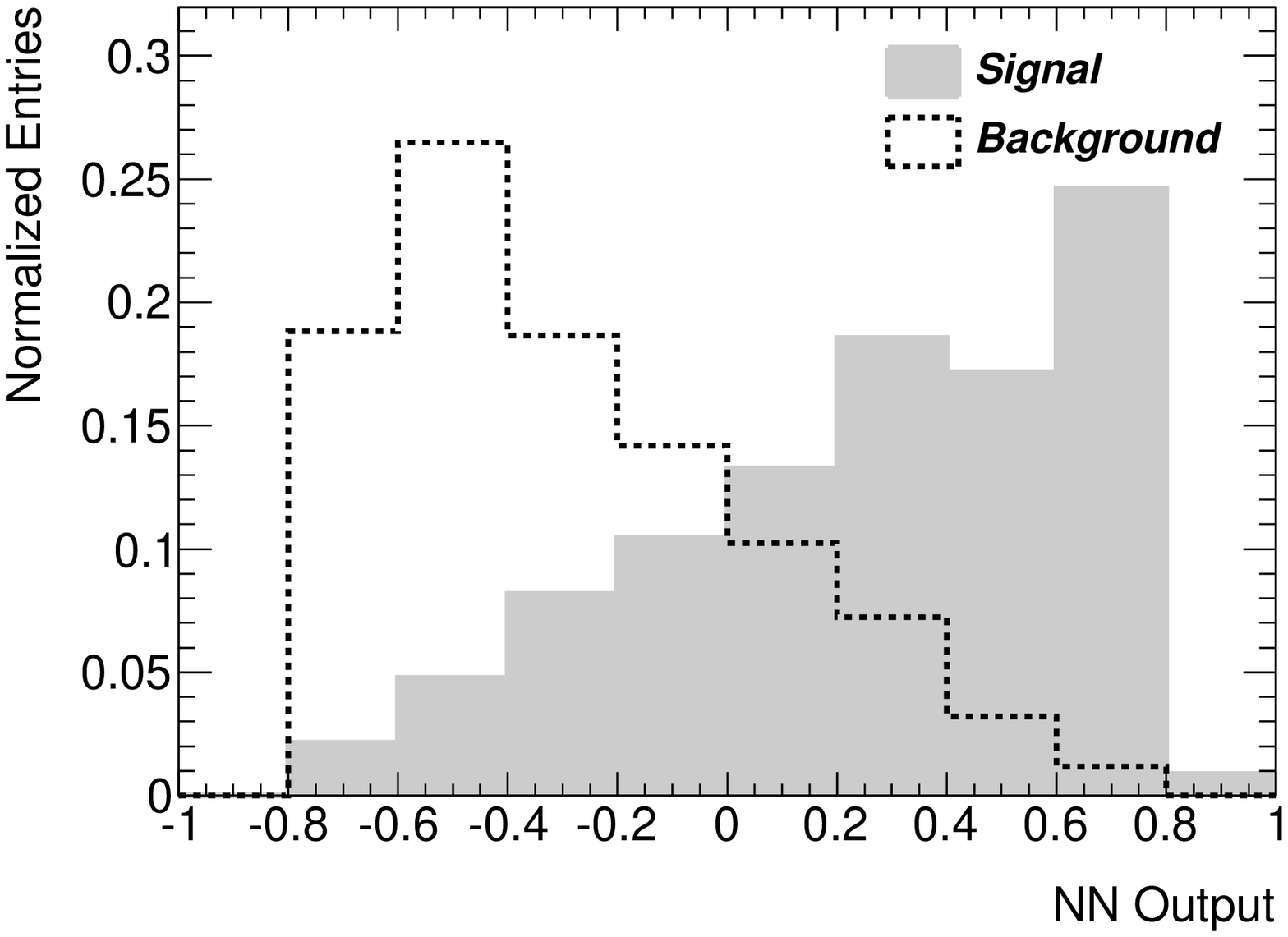}
\caption{{\bf (Left:) $H_T$ and (right:) NN discriminants for $\mu = 50$ case. They are fit in the background and signal regions respectively.
The different curves are normalized to equal area}}
\label{fig:disc}
\end{center}
\end{figure}

The data is divided into two regions, according to the number of hadronic jets and the $b$-tag multiplicity: one region is dominated by the $\ttbar$ background (2 jets, $\geq$ 1 tight b-tag), while the other is signal-enriched ($\geq$ 3 jets, $\geq$ 2 tight b-tags).  This strategy is similar to both ATLAS\cite{ATLASttH} and CMS \cite{CMSttH} analyses at 7 and 8 TeV.  Both regions are fit simultaneously using a profile likelihood fit within the MCLimit framework\cite{mclimit}. Two separate distributions are chosen for their ability to distinguish signal from background.  In the $\ttbar$ dominated region, the $H_T$ distribution (the scalar sum of the $p_T$ of jets and charged leptons) is used.  In the signal-enriched region, a Neural Network (NN) discriminant is built using the Neuro-Bayes framework~\cite{NB} to separate $\ttbar$ and electroweak background from the $\ttH$ signal. The variables that go into the Neural Network are, (a) a pseudo continuous b-tagger built from the correlation of number of loose and number of tight b-tags per event, (b) $H_T$, (c) $p_T$ of the leading jet, (d) number of jets, (e) closest distance (in $\eta, \phi$-space) between a lepton and jet, (e) closest distance between two jets. This neural net provides very good discrimination between signal and background. These two variables for the $\mu=50$ case are shown in Figure \ref{fig:disc}; the curves are normalized to equal area and serve to show the difference in shapes.

\section{Studies at 14 TeV}
\label{sect:14TeV}

Signal and background yields for integrated luminosities and pileup scenarios of 300 fb$^{-1}$ with $\mu = 50$, and 3000 fb$^{-1}$ with $\mu = 140$ are presented in Table \ref{table:Events}, for both the background-dominated and signal-enriched regions.

\begin{table}[ht!]
\centering
\begin{tabular}{|c|c|c|c|c|}
 \cline{2-5}
        \multicolumn{1}{c|}{} &
        \multicolumn{2}{|c|}{$L$ = 300 fb$^{-1}$ $\mu = 50$ pileup} &
         \multicolumn{2}{|c|}{$L$ = 3000 fb$^{-1}$ $\mu = 140$ pileup} \\
\hline
Sample & 2 jet $\geq$ 1 tag & $\geq$ 3 jet $\geq$ 2 tag & 2 jet $\geq$ 1 tag & $\geq$ 3 jet $\geq$ 2 tag \\ \hline
$\ttH$ &   46   & 770 & 248 & 7097 \\ \hline
$\ttbar$     &  894201 &  441295 & 3 937 730 & 4 793 330 \\
Electroweak  &  31916  &  2532 & 320488 & 43442 \\ \hline
$S/\sqrt{B}$ & 0.15  & 1.15 & 0.12 & 3.24 \\ \hline
\end{tabular}
\caption{{\bf Number of Events for 300 fb$^{-1}$ with $\mu = 50$ and 3000 fb$^{-1}$ with $\mu = 140$ pileup scenarios}}
\label{table:Events}
\end{table}

\subsection{Systematics}

We consider a 10\% uncertainty on the $\ttbar$ production cross section and a 5\% uncertainty on the $Z$+jets cross section, and uncertainties of +14.8\% and -18.2\% on the $\ttH$ signal cross section. These are pure normalization cross sections, and we do not consider any further theoretical uncertainties from $\ttbar$ modelling. The limited knowledge of the kinematics of $t\bar{t}b\bar{b}$ and $t\bar{t}c\bar{c}$ are currently an important limitation to the experimental search for $\ttH$, but it is assumed here that these issues will be resolved within the next few years, and will therefore not be a significant problem. We consider two sources of experimental systematic uncertainties: (a) a global Jet Energy Scale uncertainty of 5\%, and (b) a 20\% uncertainty on the $b$-tagging efficiency and light-jet mis-tag rate. In both cases the uncertainty is calculated simply as a change in jet and $b$-jet acceptance. No dependence of these sources of uncertainty on the jet dynamics is considered.

\subsection{Results}

Table~\ref{table:Results} shows the the $p$-value and expected significance assuming a Standard-Model signal strength for the two luminosity scenarios. It is important to remember that these projections are based on the dilepton channel only. The lepton+jets channel where only one of the top quarks decays to a charged lepton dominates the sensitivity of current searches at 7 and 8 TeV, and is expected to show a similar sensitivity in the conditions expected during the LHC data taking at higher center of mass energy; in addition, analyses are ongoing to study the feasibility of the
all-hadronic final state.

\begin{table}[ht!]
\centering
\begin{tabular}{|l|c|c|c|c|}
 \cline{2-5}
        \multicolumn{1}{c|}{} &
        \multicolumn{2}{|c|}{$L$ = 300 fb$^{-1}$, $\mu = 50$ pileup} &
        \multicolumn{2}{|c|}{$L$ = 3000 fb$^{-1}$, $\mu = 140$ pileup}\\
\hline
Result & Statistics Only & With Systematics & Statistics Only & With Systematics \\ \hline
$p$-value       & 0.015   & 0.017      & - & $\le 10^{-7}$ \\ \hline
Significance    &  2.4  &  2.4     & - & $\ge 5.3 $\\ \hline
\end{tabular}
\caption{{\bf Results of the Profile Likelihood Fit: $p$-value and signal significance are shown. }}

\label{table:Results}
\end{table}

\section{Conclusions and Outlook}

A study of $\ttH$ production at 14 TeV center of mass energy was performed assuming 300 fb$^{-1}$ and 3000 fb$^{-1}$ with realistic pile-up scenarios and a reasonable treatment of the more important sources of experimental systematic uncertainties. The results show that the dilepton channel alone will measure $\ttH$ with a significance of about 2.4 and $\ge 5.3$ (assuming Standard Model production rate for the Higgs) for the two luminosity
scenarios, respectively. The lepton+jets channel, which has not been addressed here, will also contribute to the experimental sensitivity.

As these studies progress, we expect that analysis techniques will improve, e.g., by using variables with better discrimination in the Neural Network,
splitting the background-dominated and signal-enriched into more jet multiplicity bins to better control systematic uncertainties, better understanding of the b-tagging
algorithms, etc., thereby improving the significance of the results.

\end{document}